\documentclass[12pt]{article}
\pdfoutput=1
\usepackage{amsmath}
\def\be{\begin{equation}}
\def\ee{\end{equation}}
\def\arr{\begin{array}{rll}}
\def\ea{\end{array}}
\def\bea{\begin{eqnarray}}
\def\eea{\end{eqnarray}}

\def\N2{$N{=}2$}

\def\>{\rangle}
\def\<{\langle}
\def\+{\dagger}
\def\={\ =\ }

\begin{document}
\renewcommand{\thefootnote}{\fnsymbol{footnote}}
\begin{titlepage}
\setcounter{page}{0}
\vskip 1cm
\begin{center}
{\LARGE\bf  Remarks on integrability of $\mathcal{N}=1$}\\
\vskip 0.4cm
{\LARGE\bf   supersymmetric Ruijsenaars--Schneider}\\
\vskip 0.4cm
{\LARGE\bf   three--body models}\\
\vskip 1cm
$
\textrm{\Large Anton Galajinsky \ }
$
\vskip 0.7cm
{\it
Tomsk Polytechnic University, 634050 Tomsk, Lenin Ave. 30, Russia} \\

\vskip 0.2cm
{e-mail: galajin@tpu.ru}
\vskip 0.5cm
\end{center}
\vskip 1cm
\begin{abstract} \noindent
Integrability of $\mathcal{N}=1$ 
supersymmetric Ruijsenaars--Schneider three--body models based upon the
potentials $W(x)=\frac{2}{x}$, $W(x)=\frac{2}{\sin{x}}$, and 
$W(x)=\frac{2}{\sinh{x}}$ is proven. The problem of constructing an
algebraically resolvable set of Grassmann--odd constants of motion is reduced to
finding a triplet of vectors such that all their scalar products can be 
expressed in terms of the original bosonic first integrals.
The supersymmetric generalizations are used to build novel 
integrable (iso)spin extensions of the respective Ruijsenaars--Schneider 
three--body systems. 

\end{abstract}


\end{titlepage}

\renewcommand{\thefootnote}{\arabic{footnote}}
\setcounter{footnote}0

\noindent
{\bf 1. Introduction}\\

Supersymmetric extensions of integrable mechanics are usually studied in connection with the
superstring theories \cite{GSW}, where they describe dynamics of zero modes, 
or in the context of microscopic description of near horizon
black hole geometries \cite{GT}. The question of how a formal supersymmetrization procedure 
affects integrability has received much less attention.
It is generally believed that a superextension of an integrable theory 
should automatically result in a larger integrable system including fermionic degrees of freedom. 
If this were the case, 
supersymmetrization would suggest an efficient way of building new integrable models. 

Because the number of fermionic degrees of freedom is in general greater than the number of
conserved supercharges at hand, integrability of a supersymmetric extension
is not a priori guaranteed. Furthermore, because fermionic integrals of motion 
are constructed from monomials in Grassmann--odd variables 
and there does not exist a division by a Grassmann--odd function \cite{BK}, 
in order to guarantee integrability in the fermionic sector one has to find
constants of motion, 
which are algebraically resolvable with 
respect to the fermionic variables. A necessary condition for this is the presence 
of a linear term in each Grassmann--odd integral of motion. 

Aiming at a better understanding of the interrelation between supersymmetry and 
integrability, in a recent work \cite{AG3}
integrability of an $\mathcal{N}=1$ 
supersymmetric extension of the Ruijsenaars--Schneider hyperbolic three--body 
model \cite{RS} was studied in detail.  In particular,
three functionally independent Grassmann--odd constants of motion were explicitly 
constructed and their algebraic resolvability was demonstrated. It was also anticipated in \cite{AG3}
that proving integrability of supersymmetric extensions for other variants in \cite{RS}
should go rather straightforward. As shown below, some of such models present a challenge.

The Ruijsenaars--Schneider systems provide interesting examples 
of integrable many--body models, equations of motion of which
involve particle velocities \cite{RS} 
\be\label{RS}
\ddot{x}_i=\sum_{j\ne i}^n {\dot x}_i {\dot x}_j W(x_{ij}), 
\ee
where $x_{ij}=x_i-x_j$, $i,j=1,\dots,n$, and $W(x)$ is one of the potentials listed below
\be\label{poten}
W(x)=\left[ \frac{2}{x}, \frac{2}{\sin{x}}, 2 \cot{x},\frac{2}{\sinh{x}}, 
2 \coth{x} \right].
\ee
Such systems enjoy symmetries, which form the Poincar\'e group
in $1+1$ dimensions, and reduce to the celebrated Calogero 
models \cite{C} in the nonrelativistic limit \cite{RS}. 
By this reason, the former are usually regarded as
the relativistic analogues of the latter.

Surprisingly enough, supersymmetric extensions of the relativistic 
counterparts of the Calogero models remain almost completely unexplored.
An $\mathcal{N}=2$ supersymmetric generalization of the quantum
trigonometric Ruijsenaars--Schneider model was constructed
in \cite{BDM} and its eigenfunctions were linked to the Macdonald superpolynomials. 
Note that the fermionic variables in \cite{BDM} and their adjoints obey  
non--standard anticommutation relations which reduce to the 
conventional ones in the non--relativistic limit only. 
$\mathcal{N}=2$ supersymmetric extensions of the rational and hyperbolic 
three--body models were built in \cite{AG1} within the Hamiltonian framework. 
The corresponding supercharges were cubic in the fermionic variables.
$n$--particle $\mathcal{N}=2$ models were suggested in \cite{KL,KKL}. 
The highest power of the fermionic degrees of freedom contributing to
the $\mathcal{N}=2$ supercharges in \cite{KL,KKL} depends on the number of particles 
at hand, making the supercharges highly nonlinear. Note that 
algebraic resolvability of constants of motion in the fermionic sector 
has not been analyzed in \cite{BDM,AG1,KL,KKL}.

The goal of this work is to extend our recent analysis in \cite{AG3}  of the integrability of
an $\mathcal{N}=1$ supersymmetric Ruijsenaars--Schneider three--body 
system based upon the potential $W(x)=\frac{2}{\sinh{x}}$ 
to other instances listed in (\ref{poten}), as well as to the Ruijsenaars--Toda model
\cite{R}. Like in \cite{AG3}, we choose to work within the Hamiltonian framework. Our approach 
includes three steps. 

Firstly, subsidiary functions $\lambda_i$ are built on the phase space parametrized by
$(x_i,p_i)$, $i=1,2,3$, $\{x_i, p_j \}=\delta_{ij}$,
which generate the potential
$W(x)$ via the Poisson bracket (no summation over repeated indices and $i \ne j$)
$\{\lambda_i,\lambda_j \}=\frac{1}{4} W(x_{ij}) \lambda_i \lambda_j$.
At the same time, they allow one to represent the Hamiltonian in the quadratic form,
$H=\lambda_i \lambda_i=I_1$. Two more constants of motion $I_2$ and $I_3$ 
available for a three--body model at hand are expressed in terms of $x_i$ and 
$\lambda_i$ as well. 

Secondly, a fermionic partner $\theta_i$ is considered
for each canonical pair $(x_i,p_i)$, which obeys the Poisson bracket
$\{\theta_i,\theta_j \}=-{\rm i} \delta_{ij}$, and a 
natural $\mathcal{N}=1$ supersymmetry charge $Q_1=\lambda_i \theta_i$ is introduced,
which generates the superextended Hamiltonian\footnote{Throughout the text, superextensions of the original bosonic 
quantities are denoted by the same letters
written in the calligraphic style.} $\mathcal{H}$ via the Poisson bracket, 
$\{Q_1,Q_1 \}=-{\rm i} \mathcal{H}$. The latter governs dynamics of the resulting
$\mathcal{N}=1$ supersymmetric Ruijsenaars--Schneider system.

Thirdly, in order to establish integrability in the fermionic sector, 
one needs to find
two more Grassmann--odd first integrals, the leading terms of which are linear in 
the Grassmann--odd variables $Q_2=\mu_i \theta_i+\dots$, $Q_3=\nu_i \theta_i+\dots$,
where dots designate terms cubic in the fermions and 
$\mu_i (x,\lambda)$, $\nu_i(x,\lambda)$ 
are specific functions to be fixed below. Because $Q_2$ and $Q_3$ are supposed to commute 
with the superextended Hamiltonian $\mathcal{H}$,
the Poisson brackets between $Q_1$, $Q_2$, and $Q_3$
should be conserved over time as well. This follows from the super Jacobi identity. 
Considering the bosonic limit of expressions contributing to the right hand 
sides of the respective brackets,
one concludes that the scalar products
$\lambda_i \mu_i$, $\lambda_i \nu_i$, $\mu_i \mu_i$, $\mu_i \nu_i$, $\nu_i \nu_i$,
should all link to the bosonic first integrals $(I_1,I_2,I_3)$ characterizing 
the original model at hand. It then remains to extract $\mu_i$ and $\nu_i$ from $(I_1,I_2,I_3)$.
To put it in other words, given a Ruijsenaars--Schneider three--body system
with three constants of motion
$(I_1,I_2,I_3)$, our approach to supersymmetrizing it consists in finding a triplet
of vectors $(\lambda_i,\mu_i,\nu_i)$, all scalar products of which are expressible in terms 
of $(I_1,I_2,I_3)$.

Surprisingly enough, as demonstrated below, 
while such a procedure works smoothly for the rational potential
$W(x)=\frac{2}{x}$, the trigonometric variant $W(x)=\frac{2}{\sin{x}}$, and its
hyperbolic analogue
$W(x)=\frac{2}{\sinh{x}}$, it unexpectedly fails for $W(x)=\cot{x}$, $W(x)=\coth{x}$, as well 
as for the Ruijsenaars--Toda case, meaning that a more sophisticated approach
of proving the algebraic resolvability in the fermionic sector of those models is needed. 
Note that in all the cases 
in which our construction proves successful, it relies upon specific rational/trigonometric 
identities (see (\ref{id1}), (\ref{TRI}), and (\ref{id3}) below),
analogues of which are missing for $W(x)=\cot{x}$, $W(x)=\coth{x}$, and
the Ruijsenaars--Toda system.

The work is organized as follows. In the next section, an integrable $\mathcal{N}=1$ 
supersymmetric extension of the Ruijsenaars--Schneider rational three--body system is 
constructed. A specific reduction is also discussed, which allows 
one to build a novel integrable (iso)spin extension of the original bosonic rational model.
In Sect. 3.2, a triplet of vectors $(\lambda_i,\mu_i,\nu_i)$ is built, which underlies an 
integrable $\mathcal{N}=1$ 
supersymmetric extension of the Ruijsenaars--Schneider trigonometric three--body 
model based upon the potential $W(x)=\frac{2}{\sin{x}}$. A respective
integrable (iso)spin extension of the original trigonometric model is proposed as well.
Difficulties in obtaining a similar triplet for $W(x)=\cot{x}$ are 
summarized in Sect. 3.2. Sect. 4.1 and 4.2 contain similar analysis of the hyperbolic analogues
based upon
$W(x)=\frac{2}{\sinh{x}}$ and $W(x)=\coth{x}$ producing similar results. In Sect. 5, it is demonstrated
that for the Ruijsenaars--Toda system it proves problematic to build a triplet of 
vectors such that all their scalar products link to first integrals of the original bosonic model.
In the concluding Sect. 6, we summarize 
our results and discuss issues deserving of further study.  

Throughout the paper summation over repeated indices is understood
unless otherwise stated. 

\vspace{0.5cm}

\noindent
{\bf 2. $\mathcal{N}=1$ supersymmetric rational model}\\

The Ruijsenaars--Schneider rational model is described by the differential equations (\ref{RS}), 
in which $W(x_{ij})=\frac{2}{x_{ij}}$,
$x_{ij}=x_i-x_j$, $i,j=1,\dots,n$, and $x_1>x_2>\dots >x_n$.
Functionally independent first integrals, which provide integrability of the system, 
read
\bea\label{FI0}
&&
I_1=\sum_{i=1}^n {\dot x}_i, 
\\[2pt]
&&
I_2=\sum_{i<j}^n {\dot x}_i {\dot x}_j {\left(x_{ij}\right)}^2, 
\nonumber\\[2pt]
&&
I_3=\sum_{i<j<k}^n {\dot x}_i {\dot x}_j {\dot x}_k {\left( x_{ij} \right)}^2
 {\left( x_{ik} \right)}^2  {\left(x_{jk}\right)}^2, 
\nonumber\\[2pt]
&&
I_4=\sum_{i<j<k<l}^n {\dot x}_i {\dot x}_j {\dot x}_k {\dot x}_l 
{\left(x_{ij} \right)}^2
{\left(x_{ik}\right)}^2  {\left(x_{il}\right)}^2 
{\left(x_{jk}\right)}^2 {\left(x_{jl}\right)}^2 
{\left(x_{kl}\right)}^2, 
\nonumber\\[2pt]
&&
\hdots
\nonumber
\eea
where $\dots$ designate higher order invariants, which can be constructed likewise. 

Our objective in this section is to construct
an $\mathcal{N}=1$ supersymmetric extension of the rational system for the three--body case
and to establish its integrability. 
To this end, it proves convenient to switch to the Hamiltonian formalism \cite{FC}, 
within which the model is represented by three mutually commuting constants of motion
\bea\label{FI1}
&&
I_1=\frac{e^{p_1}}{x_{12} x_{13}}+
\frac{e^{p_2}}{x_{12} x_{23}}+\frac{e^{p_3}}{x_{13} x_{23}},
\qquad
I_2=\frac{e^{p_1+p_2}}{x_{13} x_{23}}+\frac{e^{p_1+p_3}}{x_{12} x_{23}}+\frac{e^{p_2+p_3}}{x_{12} x_{13}},
\nonumber\\[2pt]
&&
I_3=e^{p_1+p_2+p_3},
\eea
the first of which is identified with the Hamiltonian, $H=I_1$. The Poisson bracket is chosen in the
conventional form $\{x_i,p_j \}=\delta_{ij}$. 

In order to build an $\mathcal{N}=1$ supersymmetric extension, one first introduces 
three subsidiary functions 
\begin{align}\label{sf1}
&
\lambda_1=\frac{e^{\frac{p_1}{2}}}{\sqrt{x_{12} x_{13}}}, &&
\lambda_2=\frac{e^{\frac{p_2}{2}}}{\sqrt{x_{12} x_{23}}}, &&
\lambda_3=\frac{e^{\frac{p_3}{2}}}{\sqrt{x_{13} x_{23}}},
\end{align}
which generate the potential $W(x)=\frac{2}{x}$ via the Poisson bracket
(no summation over repeated indices and $i \ne j$)
\be\label{sfr1}
\{\lambda_i,\lambda_j \}=\frac{1}{4} W(x_{ij}) \lambda_i \lambda_j. 
\ee
In terms of $\lambda_i$, the Hamiltonian takes on the quadratic form
\be
H=\lambda_i \lambda_i,
\ee
which is amenable to immediate supersymmetrization.

For most of the calculations to follow, it proves convenient to 
trade $p_i$ for $\lambda_i$, which slightly
modifies the canonical bracket (no summation over repeated indices)
\be
\{x_i, \lambda_j \}=\frac 12 \delta_{ij} \lambda_j.
\ee
The Hamiltonian equations of motion for $x_i$ and $\lambda_i$ then read
\be\label{caneq}
{\dot x}_i=\lambda_i^2, \qquad 
{\dot\lambda}_i=\frac 12 \sum_{j \ne i} W(x_{ij}) \lambda_i \lambda_j^2. 
\ee
These equations prove to maintain their form for other potentials in (\ref{poten}).
In establishing the supersymmetry algebra below, the following identity 
(no summation over repeated indices)
\be
\{ \lambda_i, x_{ij} \lambda_j \}=0
\ee
will prove useful.

As the second step, each canonical pair $(x_i,p_i)$ is accompanied by a real Grassmann--odd partner $\theta_i$,
obeying the Poisson bracket\footnote{The conventional fermionic kinetic term 
$\frac{\rm i}{2} \int dt \theta_i {\dot\theta}_i$ gives rise to the second class constraints
$p_{\theta i}-\frac{\rm i}{2} \theta_i=0$, where 
$p_{\theta i}=\frac{\partial \mathcal{L}}{\partial {\dot\theta}_i}$ is the momentum 
canonically conjugate to $\theta_i$, $\mathcal{L}=\frac{\rm i}{2} \theta_i {\dot\theta}_i$ 
is the Lagrangian density, and the right derivative with respect to the Grassmann--odd variables is used.  
Introducing the conventional Dirac bracket and eliminating $p_{\theta i}$ from the consideration 
by resolving the second class constraints,
one arrives at (\ref{DB1}).}
\be\label{DB1}
\{\theta_i,\theta_j \}=-{\rm i} \delta_{ij},
\ee 
and a natural $\mathcal{N}=1$ supersymmetry charge is introduced
\be\label{Q11}
Q_1=\lambda_i \theta_i, 
\ee
which via the Poisson bracket generates the superextended Hamiltonian
\bea\label{H1}
\{Q_1,Q_1 \}=-{\rm i} \mathcal{H}=-{\rm i} \mathcal{I}_1, \qquad \mathcal{H}=\lambda_i \lambda_i +\frac{{\rm i}}{4} W(x_{ij}) \lambda_i \lambda_j \theta_i \theta_j.
\eea

As was explained in the Introduction, in order to establish integrability in the fermionic sector, 
one needs to find
two more Grassmann--odd first integrals, the leading terms of which are linear in 
the Grassmann--odd variables
\be\label{q2q3}
Q_2=\mu_i \theta_i+\dots, \qquad Q_3=\nu_i \theta_i+\dots,
\nonumber
\ee
where $\dots$ designate terms cubic in the fermions and 
$\mu_i (x,\lambda)$, $\nu_i(x,\lambda)$ 
are specific functions to be fixed below. Because $Q_2$ and $Q_3$ are supposed to commute 
with the superextended Hamiltonian $\mathcal{H}$,
the following Poisson brackets 
\begin{align}\label{q2q2}
&
\{Q_1,Q_2 \}=-{\rm i} \lambda_i \mu_i+\dots, && \{Q_1,Q_3 \}=-{\rm i} \lambda_i \nu_i+\dots, &&
\{Q_2,Q_2 \}=-{\rm i} \mu_i \mu_i+\dots, 
\nonumber\\[2pt]
&
\{Q_2,Q_3 \}=-{\rm i} \mu_i \nu_i+\dots, &&
\{Q_3,Q_3 \}=-{\rm i} \nu_i \nu_i+\dots, &&
\nonumber
\end{align}
where $\dots$ stand for terms quadratic in the Grassmann--odd variables,
should be conserved over time as well. This follows from the super Jacobi identities. 
Considering the bosonic limit of the expressions contributing to the right hand sides,
one concludes that the scalar products
\be
\lambda_i \mu_i, \qquad \lambda_i \nu_i, \qquad \mu_i \mu_i, \qquad \mu_i \nu_i, \qquad \nu_i \nu_i,
\nonumber
\ee
should all link to the bosonic first integrals (\ref{FI1}) characterizing the model at hand.

Rewriting (\ref{FI1}) in terms of the subsidiary functions (\ref{sf1})
\be\label{FI1R}
I_1=\lambda_i \lambda_i, \qquad I_2=\frac 12 \lambda_i^2 \lambda_j^2 x_{ij}^2 ,
\qquad
I_3={\left( \frac{1}{3!} \epsilon_{ijk} \lambda_i \lambda_j 
\lambda_k x_{ij} x_{ik} x_{jk} \right)}^2,
\ee
where $\epsilon_{ijk}$ is the Levi--Civita totally antisymmetric symbol with $\epsilon_{123}=1$, one obtains a 
natural candidate for the vector $\mu_i$, which underpins $Q_2$
\be\label{VB}
I_2=\mu_i \mu_i, \qquad \mu_i=\frac 12 \epsilon_{ijk} \lambda_j \lambda_k x_{jk}, 
\qquad \lambda_i \mu_i=0, 
\ee
where the last equality holds due to the identity
\be\label{id1}
x_{12}-x_{13}+x_{23}=0.
\ee
In obtaining (\ref{VB}), the properties of the Levi--Civita symbol
\be
\epsilon_{ijk} \epsilon_{lp k}=\delta_{il} \delta_{jp}-\delta_{ip} \delta_{jl}, \qquad 
\epsilon_{ijk} \epsilon_{ljk}=2 \delta_{il}
\ee
proved useful. Then it is straightforward to verify that 
\be
Q_2=\mu_i \theta_i=\frac 12 \epsilon_{ijk} \lambda_j \lambda_k x_{jk} \theta_i,
\ee
Poisson commutes with $Q_1$ 
and, hence, it is conserved over time as a consequence of the 
super Jacobi identity involving 
the triplet $(Q_1,Q_1,Q_2)$. 
In verifying the relation $\{Q_1,Q_2 \}=0$, the identity (\ref{id1}) was used.

Computing the Poisson bracket of $Q_2$ with itself, one obtains the 
superextension of the original bosonic first integral $I_2$
\be
\{Q_2,Q_2 \}=-{\rm i} \mathcal{I}_2, \qquad
\mathcal{I}_2=\frac 12 \lambda_i^2 \lambda_j^2 x_{ij}^2+\frac{{\rm i}}{8} 
\left( \epsilon_{ijk} \theta_i \theta_j \lambda_i \lambda_j\right)
\left(\epsilon_{plk}   W(x_{pl}) x_{pk} x_{lk} \lambda_k^2  \right),
\ee
which rightly commutes with $\mathcal{H}=\mathcal{I}_1$ in (\ref{H1}). 
The third constant 
of motion $I_3$ in (\ref{FI1R}) does not acquire fermionic contributions and maintains its form after 
the superextension, $\mathcal{I}_3=I_3$, which is
a manifestation of the invariance of the resulting system under the 
translation $x'_i=x_i+a$.

The construction of $Q_3$ is less straightforward, however. It appears problematic to represent 
$I_3$ as a scalar product of a vector $\nu_i$ with itself. Another option, which 
will ultimately prove correct, 
is to take $\nu_i$ entering $Q_3$ as the vector 
product of $\lambda_i$, upon which $Q_1$ is constructed, and $\mu_i$, 
which underlies $Q_2$. A contribution to $Q_3$, which is cubic in the Grassmann--odd variables, 
is then found directly from the conservation equation $\{Q_3, \mathcal{H} \}$=0.

Yet another possibility to build $Q_3$ is to make recourse to higher order fermionic invariants 
available for the case at hand.
Taking into account the equations of motion in the fermionic sector
\be\label{FEOM}
{\dot\theta}_i =\frac{1}{2} \sum_{j \ne i}^3 W(x_{ij}) \lambda_i \lambda_j \theta_j,
\qquad W(x_{ij})=\frac{2}{x_{ij}},
\ee
one readily obtains a cubic integral of motion
\be\label{Om}
\Omega=\frac{{\rm i}}{3!} \epsilon_{ijk} \theta_i \theta_j \theta_k={\rm i} \theta_1 \theta_2 \theta_3, 
\ee
which is conserved over time as a consequence of the Grassmann--valued nature of 
the variable
$\theta_i$: $\theta_1^2=\theta_2^2=\theta_3^2=0$. The Poisson brackets of
$\Omega$ and $Q_1$, $Q_2$ can then be used to build lower order fermionic invariants
\bea
&&
\{Q_1,\Omega \}=-{\rm i } \Lambda, \qquad \Lambda=\frac{{\rm i }}{2}  \epsilon_{ijk} \lambda_i \theta_j \theta_k,
\nonumber\\[2pt]
&&
\{Q_2, \Lambda \}=Q_3, \qquad 
Q_3=x_{ij} \lambda_j^2 \lambda_i \theta_i+\frac 14 \epsilon_{ijk}
x_{ij} W(x_{jk}) \lambda_i \lambda_j \lambda_k \Omega,
\eea
the last of which is the desired third fermionic constant of motion. It is straightforward to verify that 
the leading term in $Q_3=\nu_i \theta_i+\dots$ is indeed constructed as 
the vector product of $\lambda_i$ and $\mu_j$
\be\label{Supl}
\nu_i=x_{{\hat i} j} \lambda_{\hat i} \lambda_j^2 
=\epsilon_{ijk} \lambda_j \mu_k, \qquad
\nu_i \nu_i=I_1 I_2, \qquad \nu_i \lambda_i=0, \qquad \nu_i \mu_i=0,
\ee
with $\mu_k$ defined in (\ref{VB}). In the leftmost equation in (\ref{Supl}) 
and in the text below
no summation over repeated indices carrying a hat symbol is understood. 

Thus, 
$(\lambda_i,\mu_i,\nu_i)$ do form a triplet of vectors, all scalar products of which can be expressed 
in terms of the bosonic first integrals (\ref{FI1R}). The latter fact will prove important 
in subsequent sections,
where other variants of the Ruijsenaars--Schneider model will be studied.

At this point, algebraic resolvability of the fermionic constants of motion $(Q_1,Q_2,Q_3)$ 
with respect to the variables $(\theta_1,\theta_2,\theta_3)$
can be easily established. Because the cubic term $\Omega$ is itself 
conserved over time,  the expressions for $(Q_1,Q_2,Q_3)$  can be put into
the linear algebraic form $A_{ij} \theta_j=B_i$, where $B_i$ is a specific vector function, 
which can be read off from $(Q_1,Q_2,Q_3)$, and $A_{ij}$ is the matrix involving three rows
$A_{1i}=\lambda_i$, $A_{2i}=\mu_i$, $A_{3i}=\nu_i=\epsilon_{ijk} \lambda_j \mu_k$.
Because the determinant of $A_{ij}$ is equal to the square of the area of a parallelogram formed by 
the vectors $\lambda_i$ and $\mu_i$, the matrix $A_{ij}$ is invertible and, hence, the
system of equations for $\theta_i$ is algebraically resolvable: $\theta_i={\left(A^{-1} \right)}_{ij} B_j$. 
Taking the resulting expressions 
and computing the product $\theta_1 \theta_2 \theta_3$, one can then link the higher order invariant 
$\Omega$ to $(Q_1,Q_2,Q_3)$ and $(\mathcal{I}_1,\mathcal{I}_2,\mathcal{I}_3)$.

As was mentioned in the Introduction, a formal supersymmetrization
procedure is expected to provide an efficient way of generating integrable extensions of known integrable systems. 
Concluding this section,
we discuss how the $\mathcal{N}=1$ supersymmetric model above can be used to build 
a novel 
integrable (iso)spin 
extension of the Ruijsenaars-Schneider rational three--body model.

For the case at hand, the fermionic sector 
is described by three Grassmann--odd variables
$\theta_i$, $i=1,2,3$, which obey the first order differential equations (\ref{FEOM}).
The corresponding general solution involves three Grassmann--odd constants of integration.
Denoting them by $\alpha$, $\beta$, and $\gamma$ and taking into account 
$\alpha^2=\beta^2=\gamma^2=0$, one gets the natural decompositions
\bea\label{comp}
&&
\theta_i=\alpha \varphi_{i1} + \beta \varphi_{i2}+\gamma \varphi_{i3} + 
{\rm i} \alpha \beta \gamma \varphi_{i4}, \qquad
x_i =x_{i 0} + {\rm i} \alpha\beta x_{ i 1}  +{\rm i} \alpha\gamma x_{ i 2}  
+{\rm i} \beta\gamma x_{ i 3},
\eea
where components accompanying $\alpha$, $\beta$, and $\gamma$ 
are real {\it bosonic} functions of the temporal variable $t$. 
Substituting (\ref{comp}) into the Hamiltonian equations of motion 
of the superextended system
and analyzing monomials in $\alpha$, $\beta$, $\gamma$ on both sides,
one turns them into a system of ordinary 
differential equations for usual real--valued functions. The latter provides 
an integrable extension of the original Ruijsenaars--Schneider rational model.

The resulting system is rather bulky and hard to interpret. 
A simple and tractable
extension arises 
if one focuses on a particular solution for which $\beta=\gamma=0$
\bea\label{comp1}
&&
\theta_i=\alpha \varphi_{i}, \qquad \alpha^2=0,
\eea
$\varphi_{i}$ being a real--valued bosonic function to be interpreted below as describing 
(iso)spin degrees of freedom.
In this case, all terms quadratic or cubic in the fermionic variables 
vanish owing to $\alpha^2=0$ and the $\mathcal{N}=1$ superextension above reduces to the original 
Ruijsenaars--Schneider equations (\ref{RS}), which are accompanied by the linear differential 
equations for $\varphi_i$
\be\label{ISO1}
{\dot\varphi}_i =\frac{1}{2} \sum_{j \ne i}^3 W(x_{ij}) 
\sqrt{{\dot x}_i {\dot x}_j } \varphi_j,
\qquad W(x_{ij})=\frac{2}{x_{ij}}.
\ee
The latter system inherits from its $\mathcal{N}=1$ supersymmetric progenitor three first integrals
\be\label{Extra1}
I_4=\sqrt{{\dot x}_i} \varphi_i, \qquad I_5=\frac 12 \epsilon_{ijk} \sqrt{{\dot x}_i {\dot x}_j} x_{ij} \varphi_k,
\qquad I_6=-x_{ij} {\dot x}_i \sqrt{{\dot x}_j} \varphi_j,
\ee
which descend from $(Q_1,Q_2,Q_3)$, and admits one extra constant of motion 
\be\label{sphere1}
I_7=\varphi_i \varphi_i,
\ee
which implies that $\varphi_i$ can be interpreted as internal degrees of 
freedom parametrizing a two--sphere.
To the best of our knowledge, such integrable (iso)spin extension of the 
Ruijsenaars-Schneider rational three--body model is new.

\vspace{0.5cm}

\noindent
{\bf 3. $\mathcal{N}=1$ supersymmetric trigonometric models}\\

\noindent
{\it 3.1 The case of $W(x)=\frac{2}{\sin{x}}$}\\

The potentials listed in (\ref{poten}) contain two trigonometric variants,
the first of which is described by the equations of motion (\ref{RS})
involving $W(x_{ij})=\frac{2}{\sin{x_{ij}}}$, 
with $x_{ij}=x_i-x_j$, $i,j=1,\dots,n$, $x_1>x_2>\dots >x_n$. The system is characterized by the first integrals
\bea\label{FI?}
&&
I_1=\sum_{i=1}^n {\dot x}_i, 
\\[2pt]
&&
I_2=\sum_{i<j}^n {\dot x}_i {\dot x}_j \tan^2{\left(\frac{x_{ij}}{2}\right)}, 
\nonumber\\[2pt]
&&
I_3=\sum_{i<j<k}^n {\dot x}_i {\dot x}_j {\dot x}_k \tan^2{\left(\frac{x_{ij}}{2}\right)}
 \tan^2{\left(\frac{x_{ik}}{2}\right)}  \tan^2{\left(\frac{x_{jk}}{2}\right)}, 
\nonumber\\[2pt]
&&
I_4=\sum_{i<j<k<l}^n {\dot x}_i {\dot x}_j {\dot x}_k {\dot x}_l 
\tan^2{\left(\frac{x_{ij}}{2}\right)}
\tan^2{\left(\frac{x_{ik}}{2}\right)}  \tan^2{\left(\frac{x_{il}}{2}\right)}
\tan^2{\left(\frac{x_{jk}}{2}\right)} \tan^2{\left(\frac{x_{jl}}{2}\right)}
\tan^2{\left(\frac{x_{kl}}{2}\right)}, 
\nonumber\\[2pt]
&&
\hdots
\nonumber
\eea
where $\dots$ stand for higher order invariants, which are constructed in a similar fashion. 

Like before, the Hamiltonian formulation for such a three--body model
is constructed in terms the subsidiary functions
\bea\label{sf2}
&&
\lambda_1=e^{\frac{p_1}{2}} \sqrt{ \cot{\left(\frac{x_{12}}{2}\right)}\cot{\left(\frac{x_{13}}{2}\right)}}, \qquad 
\lambda_2=e^{\frac{p_2}{2}}\sqrt{ \cot{\left(\frac{x_{12}}{2}\right)} \cot{\left(\frac{x_{23}}{2}\right)}}, 
\nonumber\\[2pt]
&&
\lambda_3=e^{\frac{p_3}{2} } 
\sqrt{\cot{\left(\frac{x_{13}}{2}\right)}\cot{\left(\frac{x_{23}}{2}\right)}},
\eea
where $p_i$ are momenta canonically conjugate to the coordinates $x_i$,  
$\{x_i,p_j \}=\delta_{ij}$, which generate the potential $W(x)=\frac{2}{\sin{x}}$
via the Poisson bracket (no summation over repeated indices and $i \ne j$)
\be
\{\lambda_i,\lambda_j \}=\frac{1}{4} W(x_{ij}) \lambda_i \lambda_j, \qquad 
W(x_{ij})=\frac{2}{\sin{x_{ij}}}.
\ee
In terms of $\lambda_i$, three functionally independent integrals of motion 
in involution take on the form
\bea\label{FI2}
&&
I_1=\lambda_i \lambda_i, \qquad I_2=\frac 12 \lambda_i^2 \lambda_j^2 \tan^2{\left(\frac{x_{ij}}{2}\right)} ,
\nonumber\\[2pt]
&&
I_3={\left( \frac{1}{3!} \epsilon_{ijk} \lambda_i \lambda_j 
\lambda_k \tan{\left(\frac{x_{ij}}{2}\right)} \tan{\left(\frac{x_{ik}}{2}\right)} 
\tan{\left(\frac{x_{jk}}{2}\right)} \right)}^2,
\eea
the first of which is identified with the Hamiltonian $I_1=H$. The Hamiltonian 
equations of motion read as in (\ref{caneq}), but this time they involve 
$W(x_{ij})=\frac{2}{\sin{x_{ij}}}$.

The structure of invariants (\ref{FI2}) suggests introducing two vectors 
\be
\mu_i=\frac 12 \epsilon_{ijk} \lambda_j \lambda_k 
\tan{\left(\frac{x_{jk}}{2}\right)}, \qquad 
\nu_i=\lambda_{\hat i} \lambda_j^2 \tan{\left(\frac{x_{{\hat i} j}}{2}\right)}
=\epsilon_{ijk} \lambda_j \mu_k,
\ee
which accompany $\lambda_i$ in (\ref{sf2}). 
A remarkable property of the triplet
$(\lambda_i,\mu_i,\nu_i)$ is that all their scalar products can be expressed in terms 
of the first integrals (\ref{FI2})
\begin{align}
&
\lambda_i \lambda_i=I_1, && \mu_i \mu_i=I_2, && \nu_i \nu_i=I_1 I_2-I_3,
\nonumber\\[2pt]
&
\lambda_i \mu_i=-\sqrt{I_3}, && \lambda_i \nu_i=0, && \mu_i \nu_i=0.
\end{align}
When computing $\lambda_i \mu_i$ the trigonometric identity
\be\label{TRI}
\tan{\left(\frac{x_{12}}{2}\right)}-\tan{\left(\frac{x_{13}}{2}\right)}+
\tan{\left(\frac{x_{23}}{2}\right)}=
-\tan{\left(\frac{x_{12}}{2}\right)} \tan{\left(\frac{x_{13}}{2}\right)} 
\tan{\left(\frac{x_{23}}{2}\right)}
\ee
was used. The latter is an analogue of (\ref{id1}), which underpins the rational case.

An integrable $\mathcal{N}=1$ supersymmetric extension of the trigonometric model 
at hand is 
built upon the triplet $(\lambda_i,\mu_i,\nu_i)$ in a remarkably succinct 
way\footnote{In the unfolded form, the Poisson bracket entering (\ref{QV2}) 
reads $\{\lambda_i,\mu_i \}=
-\frac 14 \epsilon_{pjk} \tan{\left(\frac{x_{pj}}{2} \right)} W(x_{jk}) \lambda_p \lambda_j \lambda_k$.}
\be\label{QV2}
Q_1=\lambda_i \theta_i, \qquad Q_2=\mu_i \theta_i, \qquad
Q_3=\nu_i \theta_i-\{\lambda_i,\mu_i \} \Omega, 
\ee
where $\theta_i$ are the fermionic degrees of freedom obeying $\{\theta_i,\theta_j \}=-{\rm i} \delta_{ij}$
and $\Omega=\frac{{\rm i}}{3!} \epsilon_{ijk} \theta_i \theta_j \theta_k$ is the cubic 
invariant similar to that used in the previous section. Superextensions of the original bosonic first integrals 
(\ref{FI2}) are found by computing the Poisson brackets
\be
\{Q_1,Q_1 \}=-{\rm i} \mathcal{I}_1, \qquad 
\{Q_2,Q_2 \}=-{\rm i} \mathcal{I}_2, \qquad
\{Q_1,Q_2 \}={\rm i} \sqrt{\mathcal{I}_3}, 
\ee
which yield
\bea\label{SE2}
\mathcal{I}_1=\mathcal{H}=\lambda_i \lambda_i+\frac{{\rm i}}{4} W(x_{ij}) 
\lambda_i \lambda_j \theta_i \theta_j,
\qquad
\mathcal{I}_2=\mu_i \mu_i-\frac{{\rm i}}{4} W(x_{ij}) \mu_i \mu_j \theta_i \theta_j,
\qquad \mathcal{I}_3=I_3.
\eea
Like before, $I_3$ in (\ref{FI2}) does not acquire fermionic corrections in the process 
of supersymmetrization, 
which reflects the invariance of the system under the translation $x'_i=x_i+a$.
In obtaining (\ref{SE2}) the following relations 
(no summation over repeated indices and $i \ne j$)
\be
\{ \lambda_i, \tan{\left(\frac{x_{ij}}{2}\right)} \lambda_j \}=0, \qquad
\{\mu_i,\mu_j \}=-\frac{1}{4} W(x_{ij}) \mu_i \mu_j, \qquad
\{\lambda_i,\mu_j \}=\frac 14 \delta_{ij} \lambda_j \mu_j \sum_{k \ne i} W(x_{ik})
\nonumber
\ee
proved useful. The algebraic resolvability of the first integrals (\ref{QV2}) with respect to
the variables $\theta_i$ is established by 
repeating the argument in the preceding section.

Concluding this section, we display an
integrable (iso)spin extension of the trigonometric three--body model based upon
the potential $W(x)=\frac{2}{\sin{x}}$.
It is built following the recipe in the preceding section. The (iso)spin degrees 
of freedom obey the differential equations
\be\label{ISO2}
{\dot\varphi}_i =\frac{1}{2} \sum_{j \ne i}^3 W(x_{ij}) \sqrt{{\dot x}_i {\dot x}_j } \varphi_j,
\qquad W(x_{ij})=\frac{2}{\sin{x_{ij}}},
\ee
which are characterized by three constants of motion 
\be\label{Extra2}
I_4=\sqrt{{\dot x}_i} \varphi_i, \quad I_5=\frac 12 \epsilon_{ijk} \sqrt{{\dot x}_i {\dot x}_j} 
\tan{\left(\frac{x_{ij}}{2}\right)} \varphi_k,
\quad
I_6=-\tan{\left(\frac{x_{ij}}{2}\right)} {\dot x}_i \sqrt{{\dot x}_j} \varphi_j,
\ee
originating from the supercharges $(Q_1,Q_2,Q_3)$ in (\ref{QV2}). The sector also admits
an extra integral of motion 
$I_7=\varphi_i \varphi_i$ describing the geometry of the subspace parametrized by the
internal degrees of freedom. 
To the best of our knowledge, such an
integrable extension of the Ruijsenaars-Schneider trigonometric three--body model
is new.

\vspace{0.5cm}

\noindent
{\it 3.2 The case of $W(x)=2\cot{x}$}\\

The second trigonometric model builds upon the potential $W(x)=2 \cot{x}$ and
the set of functionally independent first integrals 
\bea\label{FI21}
&&
I_1=\sum_{i=1}^n {\dot x}_i, 
\\[2pt]
&&
I_2=\sum_{i<j}^n {\dot x}_i {\dot x}_j \sin^2{x_{ij}}, 
\nonumber\\[2pt]
&&
I_3=\sum_{i<j<k}^n {\dot x}_i {\dot x}_j {\dot x}_k \sin^2 {x_{ij} } 
\sin^2 {x_{ik} } \sin^2 {x_{jk}}, 
\nonumber\\[2pt]
&&
I_4=\sum_{i<j<k<s}^n {\dot x}_i {\dot x}_j {\dot x}_k {\dot x}_s 
\sin^2 {x_{ij} } 
\sin^2 {x_{ik}} \sin^2 {x_{is}} 
\sin^2 {x_{jk}} \sin^2 {x_{js}} 
\sin^2 {x_{ks}}, 
\nonumber\\[2pt]
&&
\hdots
\nonumber
\eea
where $\dots$ denote higher order invariants of a similar structure.

An $\mathcal{N}=1$ supersymmetric extension of the
three--body model at hand is constructed following the
general pattern above. It suffices to consider
three subsidiary functions
\be
\lambda_1=\frac{e^{\frac{p_1}{2}}}{\sqrt{\sin{x_{12} } \sin{x_{13} }}}, 
\qquad 
\lambda_2=\frac{e^{\frac{p_2}{2}}}{\sqrt{\sin{x_{12}} \sin{x_{23} }}}, \qquad
\lambda_3=\frac{e^{\frac{p_3}{2}}}{\sqrt{\sin{x_{13} } \sin{x_{23} }}},  
\ee
which generate the potential $W(x)=2 \cot{x}$ via the Poisson bracket (no summation over repeated indices and $i \ne j$)
\be
\{\lambda_i, \lambda_j \}=\frac 14 W(x_{ij}) \lambda_i \lambda_j,
\qquad W(x_{ij})=2 \cot{x_{ij}}.
\ee
Then one introduces a superpartner $\theta_i$ for each canonical pair $(x_i,p_i)$, and 
finally builds the linear supercharge $Q_1=\lambda_i \theta_i$. The latter generates the superextended 
Hamiltonian via the Poisson bracket, $\{Q_1,Q_1 \}=-{\rm i} \mathcal{H}$.

In order to obtain two more supercharges, one rewrites three available
first integrals in terms of $\lambda_i$
\be\label{FFI2}
I_1=\lambda_i \lambda_i, \qquad 
I_2=\frac 12 \lambda_i^2 \lambda_j^2 \sin^2{x_{ij}} ,
\qquad
I_3={\left( \frac{1}{3!} \epsilon_{ijk} \lambda_i \lambda_j 
\lambda_k \sin{x_{ij}} \sin{x_{ik}} 
\sin{x_{jk}} \right)}^2,
\ee
and then tries to extract from them two more vectors $\mu_i$ and $\nu_i$ suitable for building $Q_2$ and $Q_3$, respectively.
At this point, one reveals a problem, however. Using $I_2$ in order to construct $\mu_i$
\be
\mu_i=\frac 12 \epsilon_{ijk} \lambda_j \lambda_k 
\sin{x_{jk}}, \qquad \mu_i \mu_i=I_2,
\ee
just like we did in our previous examples,
one immediately finds that the scalar product of $\mu_i$ and $\lambda_i$ is not conserved 
over time\footnote{Of course, one can reshuffle the components of $\mu_i$ without changing
$I_2=\mu_i \mu_i$. Unfortunately, this arbitrariness does not help to improve ${\left(\lambda_i \mu_i \right)}^{\cdot} \ne 0$.}
\be
\lambda_i \mu_i=\frac{\sqrt{I_3}}{2 \cos{\left(\frac{x_{12}}{2} \right)} 
\cos{\left(\frac{x_{13}}{2} \right)} \cos{\left(\frac{x_{23}}{2} \right)}}, 
\qquad {\left(\lambda_i \mu_i \right)}^{\cdot} \ne 0.
\ee
 Note that in two previous cases the equalities
$\lambda_i \mu_i=0$ and $\lambda_i \mu_i=-\sqrt{I_3}$ held due to the identities 
(\ref{id1}) and (\ref{TRI}), respectively, which link to the specific form of the potential 
$W(x)$ for each respective case. For the model under consideration
\be
\sin{x_{12}}-\sin{x_{13}}+\sin{x_{23}} \ne \sin{x_{12}} \sin{x_{13}} \sin{x_{23}}
\ee
and, hence, $\lambda_i \mu_i$ fails to be proportional to $\sqrt{I_3}$.

One could try to treat $\sqrt{I_3}$ as a scalar product of $\lambda_i$ and
$\nu_i=-\frac 12 \epsilon_{ijk} \cos{x_{ij}} 
\cos{x_{ik}}\sin{x_{jk}} \lambda_j \lambda_k $, 
which would rely upon the 
trigonometric identity
\be
-\sin{x_{12}} \cos{x_{13}} \cos{x_{23}} + \sin{x_{13}} \cos{x_{12}} \cos{x_{23}} - 
 \sin{x_{23}} \cos{x_{12}} \cos{x_{13}}=\sin{x_{12}} \sin{x_{13}} \sin{x_{23}}.
\nonumber
\ee
Yet, at the next step one would
immediately find that $\nu_i \nu_i$ is not conserved over time. 
In a similar fashion, one could try to regard $I_2$ in (\ref{FFI2}) as 
the scalar product of $\lambda_i$ and
$\nu_i=\frac 12 \lambda_i \lambda_j^2 \sin^2{x_{ij}}$, which would
again result in the nonconservation of $\nu_i \nu_i$ over time.

Thus, despite our anticipation in \cite{AG3} that proving 
integrability in the fermionic sector should go rather straightforward for each
$\mathcal{N}=1$ supersymmetric variant of the Ruijsennars--Schneider three--body
system, the trigonometric model above presents a challenge. In the next sections, we shall 
see more examples of such a kind.

\vspace{0.5cm}

\noindent
{\bf 4. $\mathcal{N}=1$ supersymmetric hyperbolic models}\\

\noindent
{\it 4.1 The case of $W(x)=\frac{2}{\sinh{x}}$}\\

The trigonometric models above have two hyperbolic analogues, which we discuss in 
this section.\footnote{Note that the hyperbolic versions follow from 
the trigonometric models by the formal substitution $x_i\to {\rm i} x_i$. 
For completeness of the presentation, we briefly discuss them in this section.}
The first variant is based upon $W(x)=\frac{2}{\sinh{x}}$ and it
was studied in our recent work \cite{AG3}. 
Referring the reader to \cite{AG3} for more details,
we proceed directly to the subsidiary vector $\lambda_i$
\bea\label{sf3}
&&
\lambda_1=e^{\frac{p_1}{2}} \sqrt{ \coth{\left(\frac{x_{12}}{2}\right)}\coth{\left(\frac{x_{13}}{2}\right)}}, \qquad 
\lambda_2=e^{\frac{p_2}{2}}\sqrt{ \coth{\left(\frac{x_{12}}{2}\right)} \coth{\left(\frac{x_{23}}{2}\right)}}, 
\nonumber\\[2pt]
&&
\lambda_3=e^{\frac{p_3}{2} } \sqrt{\coth{\left(\frac{x_{13}}{2}\right)}\coth{\left(\frac{x_{23}}{2}\right)}},
\eea
and its two companions
\bea
&&
\mu_i=\frac 12 \epsilon_{ijk} \lambda_j \lambda_k 
\tanh{\left(\frac{x_{jk}}{2}\right)}, \qquad 
\nu_i=\lambda_{\hat i} \lambda_j^2 \tanh{\left(\frac{x_{{\hat i} j}}{2}\right)}
=\epsilon_{ijk} \lambda_j \mu_k.
\eea

In accord with our analysis above, in order to establish integrability in the fermionic sector of the corresponding
$\mathcal{N}=1$ supersymmetric extension, it suffices to verify that all scalar products 
between $(\lambda_i,\mu_i,\nu_i)$ can be expressed in terms of the
first integrals characterizing the case
\bea\label{FI3}
&&
I_1=\lambda_i \lambda_i, \qquad 
I_2=\frac 12 \lambda_i^2 \lambda_j^2 \tanh^2{\left(\frac{x_{ij}}{2}\right)} ,
\nonumber\\[2pt]
&&
I_3={\left( \frac{1}{3!} \epsilon_{ijk} \lambda_i \lambda_j 
\lambda_k \tanh{\left(\frac{x_{ij}}{2}\right)} \tanh{\left(\frac{x_{ik}}{2}\right)} 
\tanh{\left(\frac{x_{jk}}{2}\right)} \right)}^2.
\eea
An easy calculation yields
\begin{align}
&
\lambda_i \lambda_i=I_1, && \mu_i \mu_i=I_2, && \nu_i \nu_i=I_1 I_2-I_3,
\nonumber\\[2pt]
&
\lambda_i \mu_i=\sqrt{I_3}, && \lambda_i \nu_i=0, && \mu_i \nu_i=0,
\end{align}
meaning that $(\lambda_i,\mu_i,\nu_i)$ do pass the test.
Note that, like in our integrable examples above, the equality $\lambda_i \mu_i=\sqrt{I_3}$ appeals to the specific identity
\be\label{id3}
\tanh{\left(\frac{x_{12}}{2}\right)}-\tanh{\left(\frac{x_{13}}{2}\right)}+
\tanh{\left(\frac{x_{23}}{2}\right)}=
\tanh{\left(\frac{x_{12}}{2}\right)} \tanh{\left(\frac{x_{13}}{2}\right)} 
\tanh{\left(\frac{x_{23}}{2}\right)},
\ee
which holds for the hyperbolic functions at hand. The construction of
an integrable $\mathcal{N}=1$ supersymmetric extension and the respective (iso)spin reduction 
is then straightforward \cite{AG3}.

\vspace{0.5cm}

\noindent
{\it 4.2 The case of $W(x)=2 \coth{x}$}\\

The second hyperbolic model builds upon the potential $W(x)=2 \coth{x}$ and
the set of functionally independent integrals of motion 
\bea\label{FI4}
&&
I_1=\sum_{i=1}^n {\dot x}_i, 
\\[2pt]
&&
I_2=\sum_{i<j}^n {\dot x}_i {\dot x}_j \sinh^2 {\left(x_{ij}\right)}, 
\nonumber\\[2pt]
&&
I_3=\sum_{i<j<k}^n {\dot x}_i {\dot x}_j {\dot x}_k \sinh^2 {\left( x_{ij} \right)} 
\sinh^2 {\left( x_{ik} \right)} \sinh^2 {\left(x_{jk}\right)}, 
\nonumber\\[2pt]
&&
I_4=\sum_{i<j<k<s}^n {\dot x}_i {\dot x}_j {\dot x}_k {\dot x}_s 
\sinh^2 {\left(x_{ij} \right)} 
\sinh^2 {\left(x_{ik}\right)} \sinh^2 {\left(x_{is}\right)} 
\sinh^2 {\left(x_{jk}\right)} \sinh^2 {\left(x_{js}\right)} 
\sinh^2 {\left(x_{ks}\right)}, 
\nonumber\\[2pt]
&&
\hdots
\nonumber
\eea
where $\dots$ denote higher order invariants, which are constructed likewise.  

Focusing on the three--body case, introducing momenta
$p_i$ canonically conjugate to the configuration space variables $x_i$, the conventional Poisson bracket 
$\{x_i,p_j \}=\delta_{ij}$, and the Hamiltonian function
\be\label{HAM}
H=\frac{e^{p_1}}{\sinh{\left(x_{12} \right)} \sinh{\left(x_{13} \right)}} +\frac{e^{p_2}}{\sinh{\left(x_{12} \right)} \sinh{\left(x_{23} \right)}} 
+\frac{e^{p_3}}{\sinh{\left(x_{13} \right)} \sinh{\left(x_{23} \right)}}=I_1, 
\ee
one can represent the system in the Hamiltonian form. Two extra integrals of motion read
\be\label{I2I3}
I_2=\frac{e^{p_1+p_2}}{\sinh{\left(x_{13} \right)} \sinh{\left(x_{23} \right)}}+
\frac{e^{p_1+p_3}}{\sinh{\left(x_{12} \right)} \sinh{\left(x_{23} \right)}}
+\frac{e^{p_2+p_3}}{\sinh{\left(x_{12} \right)} \sinh{\left(x_{13} \right)}}, \quad I_3=e^{p_1+p_2+p_3}.
\ee
It is straightforward to verify that $(I_1,I_2,I_3)$ are functionally independent and mutually commuting,
which guarantees the Liouville integrability.

Like in all our examples above, in order to construct an $\mathcal{N}=1$ supersymmetric extension, it suffices
to build three subsidiary functions 
\be
\lambda_1=\frac{e^{\frac{p_1}{2}}}{\sqrt{\sinh{\left(x_{12} \right)} \sinh{\left(x_{13} \right)}}}, 
\quad 
\lambda_2=\frac{e^{\frac{p_2}{2}}}{\sqrt{\sinh{\left(x_{12} \right)} \sinh{\left(x_{23} \right)}}},
\quad
\lambda_3=\frac{e^{\frac{p_3}{2}}}{\sqrt{\sinh{\left(x_{13} \right)} \sinh{\left(x_{23} \right)}}},  
\nonumber
\ee
which obey the Poisson bracket (no summation over repeated indices and $i \ne j$)
\be
\{\lambda_i, \lambda_j \}=\frac 14 W(x_{ij}) \lambda_i \lambda_j, \qquad W(x_{ij})=2 \coth(x_{ij}),
\ee
then one introduces the superpartners $\theta_i$ of $(x_i,p_i)$ and 
builds the supersymmetry generator $Q_1=\lambda_i \theta_i$.
The latter generates the superextended 
Hamiltonian via the Poisson bracket, $\{Q_1,Q_1 \}=-{\rm i} \mathcal{H}$.

In the search for two more Grassmann--odd constants of motion
$Q_2=\mu_i \theta_i+\dots$ and $Q_3=\nu_i \theta_i+\dots$, one rewrites (\ref{HAM}), (\ref{I2I3})
in terms of $\lambda_i$
\be\label{FFI3}
I_1=\lambda_i \lambda_i, \qquad 
I_2=\frac 12 \lambda_i^2 \lambda_j^2 \sinh^2{x_{ij}} ,
\qquad
I_3={\left( \frac{1}{3!} \epsilon_{ijk} \lambda_i \lambda_j 
\lambda_k \sinh{x_{ij}} \sinh{x_{ik}} 
\sinh{x_{jk}} \right)}^2,
\nonumber
\ee
and then considers a feasible candidate for $\mu_i$ 
\be
\mu_i=\frac 12 \epsilon_{ijk} \lambda_j \lambda_k 
\sinh{x_{jk}}, \qquad \mu_i \mu_i=I_2.
\ee
Yet, although $\mu_i \mu_i$ is conserved over time, $\lambda_i \mu_i$ is not
\be
\lambda_i \mu_i=-\frac{\sqrt{I_3}}{2 \cosh{\left(\frac{x_{12}}{2} \right)} 
\cosh{\left(\frac{x_{13}}{2} \right)} \cosh{\left(\frac{x_{23}}{2} \right)}}, 
\qquad {\left(\lambda_i \mu_i \right)}^{\cdot} \ne 0.
\ee
The latter fact links to the inequality
\be
\sinh{x_{12}}-\sinh{x_{13}}+\sinh{x_{23}} \ne \sinh{x_{12}} \sinh{x_{13}} \sinh{x_{23}},
\ee
which prevents $\lambda_i \mu_i$ from being proportional to 
$\sqrt{I_3}$.\footnote{One could try to regard $I_2$ or $\sqrt{I_3}$ 
as the scalar product of $\lambda_i$ with
$\nu_i=\frac 12 \lambda_i \lambda_j^2 \sinh^2{x_{ij}}$ or $\nu_i=\frac{1}{3!} 
\epsilon_{ijk}  \lambda_j 
\lambda_k \sinh{x_{ij}} \sinh{x_{ik}} 
\sinh{x_{jk}}$, respectively. Yet,  $\nu_i \nu_i$ would not be conserved over time.}

Thus, similarly to its
trigonometric partner discussed in Sect. 3.2., one faces a problem in establishing
integrability in the fermionic sector of the $\mathcal{N}=1$
supersymmetric hyperbolic system at hand, which calls for a more sophisticated analysis.

\vspace{0.5cm}

\noindent
{\bf 5. $\mathcal{N}=1$ supersymmetric Ruijsenaars--Toda model}\\

Our next example is the Ruijsenaars--Toda periodic lattice, which is described by the equations 
of motion \cite{R}
\be\label{TL}
\ddot{x}_i={\dot x}_{i+1} {\dot x}_i W(x_{i+1}-x_i)-{\dot x}_i {\dot x}_{i-1} W(x_i-x_{i-1}), 
\qquad W(x-y)=\frac{g^2 e^{x-y}}{1+g^2 e^{x-y}},
\ee
where $i=1,\dots,N$, $g$ is a coupling constant. The boundary conditions
\be\label{BC}
x_0=x_N, \qquad x_{N+1}=x_1
\ee
are assumed to hold.

Introducing momenta $p_i$ canonically conjugate to the configuration space variables $x_i$ and 
the conventional Poisson bracket, $\{x_i,p_j\}=\delta_{i,j}$, one finds that 
the boundary conditions (\ref{BC}) imply 
\be\label{sr}
\{x_{i+1},p_j \}=\delta_{i+1,j}+\delta_{i,N}\delta_{j,1}, \qquad 
\{x_{i-1},p_j \}=\delta_{i-1,j}+\delta_{i,1}\delta_{j,N}.
\ee
The latter relations can be used to verify that the positive definite Hamiltonian (no sum with respect to $i$ in the second relation)
\be
H=e^{p_i}\left(1+g^2 e^{x_{i+1}-x_i}\right)=\lambda_i \lambda_i, 
\qquad \lambda_i=e^{\frac{p_i}{2}}\sqrt{1+g^2 e^{x_{i+1}-x_i}}
\ee
does put (\ref{TL}) into the Hamiltonian form. The subsidiary functions $\lambda_i$ obey the structure relations \cite{AG2}
\be\label{All}
\{\lambda_i,\lambda_j\}=\frac 14 \lambda_i \lambda_j \left( W(x_{i+1}-x_i)[\delta_{i+1,j}+\delta_{i,N}\delta_{j,1}]- W(x_{j+1}-x_j)[\delta_{i,j+1}+\delta_{i,1}\delta_{j,N}]\right).
\ee

Focusing on the three--body case and introducing a fermionic partner $\theta_i$ for each bosonic canonical pair $(x_i,p_i)$, 
one immediately obtains an $\mathcal{N}=1$ supersymmetric extension of the model at hand, which is govern 
by the supersymmetry charge $Q_1=\lambda_i \theta_i$. The latter generates the superextended 
Hamiltonian via the Poisson bracket, $\{Q_1,Q_1 \}=-{\rm i} \mathcal{H}$.

Representing three mutually commuting first integrals in terms of $\lambda_i$
\bea
&&
I_1=H=\lambda_1^2+\lambda_2^2+\lambda_3^2, \qquad I_2=\frac{\lambda_1^2 \lambda_2^2}{1+g^2 e^{x_2-x_1}}+
\frac{\lambda_1^2 \lambda_3^2}{1+g^2 e^{x_1-x_3}}+\frac{\lambda_2^2 \lambda_3^2}{1+g^2 e^{x_3-x_2}},
\nonumber\\[2pt]
&&
I_3=\frac{\lambda_1^2 \lambda_2^2 \lambda_3^2}{\left(1+g^2 e^{x_2-x_1}\right)\left(1+g^2 e^{x_3-x_2}\right)
\left(1+g^2 e^{x_1-x_3}\right)},
\eea
one can then verify that it proves problematic to construct a vector $\mu_i$, $I_2=\mu_i \mu_i$, 
such that $\lambda_i \mu_i$ is conserved over time. 
Thus, similarly to the examples in Sect. 3.2 and 4.2, 
the system does not pass our simple integrability test and a more sophisticated analysis is needed.

\vspace{0.5cm}

\noindent
{\bf 6. Conclusion}\\

To summarize, in this work integrability of $\mathcal{N}=1$ 
supersymmetric Ruijsenaars--Schneider three--body models based upon the
potentials $W(x)=\frac{2}{x}$, $W(x)=\frac{2}{\sin{x}}$, and 
$W(x)=\frac{2}{\sinh{x}}$ was proven. The problem of constructing an
algebraically resolvable set of Grassmann--odd constants of motion was reduced to
building a triplet of vectors such that all their scalar products are
expressible in terms of the original bosonic first integrals.
The supersymmetric generalizations were then used to build novel 
integrable (iso)spin extensions of the respective Ruijsenaars--Schneider 
three--body systems. 

In cases where our method succeeded, it relied upon specific
rational/trigonometric identities ((\ref{id1}), (\ref{TRI}), and (\ref{id3})). The absence of similar identities presented an obstacle for
establishing integrability of the $\mathcal{N}=1$ supersymmetric three--body systems relying upon
$W(x)=2\cot{x}$, $W(x)=2\coth{x}$, and the Ruijsenaars--Toda potential.
It is important to understand whether this is a purely technical problem or 
something more fundamental lies behind it. 

Another question deserving of further study is the construction of a Lax pair in the fermionic 
sector of the $\mathcal{N}=1$ supersymmetric systems constructed in this work. Within the Lax
formalism, constants of motion link to $\mbox{Tr} L^n$, $n=1,2,\dots$, where $L$ is the Lax matrix. 
Given three Grassmann--odd
integrals of motion $(Q_1,Q_2,Q_3)$, the leading terms of which are linear in the fermionic 
variables, the first trace is usually related to the supersymmetry charge, $\mbox{Tr} L=Q_1$.
It is interesting to study whether the higher traces $\mbox{Tr} L^n$, with $n>1$, factorize as the products 
of $(Q_1,Q_2,Q_3)$, or an alternative Lax pair can be build for each member of the triplet $(Q_1,Q_2,Q_3)$. A related 
issue is how the Lax pairs acting in the bosonic and fermionic sectors transform 
under the $\mathcal{N}=1$ supersymmetry transformations.

An extension of the present analysis to the case of more than three interacting (super)particles is 
worth studying as well. It is intriguing to see whether the construction of $n$ supercharges 
can be reduced to purely algebraic problem of building $n$ vectors, all scalar products 
of which link to $n$ first integrals characterizing the original bosonic model. Note, however, that examples 
are known in the literature, when integrability essentially depends on the number of particles.
The classic instance is the system of $n$ point vortices on a plane, which is integrable for
$n=1,2,3$ only \cite{Z}.

From the Lie--theoretic standpoint, the Ruijsenaars--Schneider $n$--body
systems are built upon root vectors of the 
simple Lie algebra $\mathcal{A}_{n-1}$. Integrable generalizations, which link
to root vectors
of other classical Lie algebras, were proposed in \cite{D}. 
The construction of supersymmetric extensions of 
the models in \cite{D} and the study of their integrability is 
an interesting avenue to explore.

Apart from root vectors underlying the classical Lie algebras, one can also consider 
deformed root systems (see reviews \cite{Ch,LST} and references therein). For example,
a deformation of the $\mathcal{A}_2$ root system amounts to keeping $x_{12}$ intact and 
changing $x_{13}$ and $x_{23}$ as follows
\be
x_{13} \to x_1-\sqrt{m} x_3, \qquad x_{23} \to x_2-\sqrt{m} x_3,
\nonumber
\ee
where $m$ is a real deformation parameter. Interestingly enough, integrability of the 
deformed models
of the Calogero type relies upon specific rational/trigonometric identities which look 
akin to those revealed in this paper.\footnote{The author thanks an anonymous JHEP 
reviewer for drawing his attention to this fact as well as for revealing the review \cite{Ch}.} Notably, the relations
(\ref{id1}), (\ref{TRI}), and (\ref{id3}) continue to hold true after the deformation
was implemented. To the best of our knowledge, 
integrability of the Ruijsenaars--Schneider--type models involving
the deformed $\mathcal{A}_{n-1}$ root system has not yet been established. 
A detailed analysis of this issue as well as the study of possible supersymmetric extensions
represent interesting open problems to tackle.

A generalization of the present research to encompass various supersymmetric extensions 
of the Calogero model is worth studying as well. In the latter regard, the similarity 
transformation in \cite{GLP} might prove helpful.

\vspace{0.5cm}

\noindent{\bf Acknowledgements}\\

\noindent
The author thanks T. Snegirev for the collaboration at an earlier stage of this project.
This work is supported in part by RF Ministry of Science and Higher Education under the 
assignment FSWW--2023--0003.

\end{document}